\documentclass[12pt,draftcls,onecolumn]{IEEEtran}

\usepackage{amsmath}
\usepackage{amssymb}
\usepackage{mathptmx}       % selects Times Roman as basic font
\usepackage{helvet}         % selects Helvetica as sans-serif font
\usepackage{courier}        % selects Courier as typewriter font
\usepackage{type1cm}        % activate if the above 3 fonts are
\usepackage{xcolor} % colori
\definecolor{verde}{rgb}{0.13,0.54,0.13}
\usepackage{makeidx}         % allows index generation
\usepackage{graphicx}        % standard LaTeX graphics tool
\usepackage{multicol}        % used for the two-column index
\usepackage[bottom]{footmisc}% places footnotes at page bottom
\usepackage{listings}

\newtheorem{definition}{Definition}

\newcommand {\be}{\begin{equation}}
\newcommand {\ee}{\end{equation}}

\begin {document}

\title{A Boolean Control Network Approach to the Formal Verification of Feedback Context-Aware Pervasive Systems}

\author{Fabio A. Schreiber  and Maria Elena Valcher
\footnote{F. Schreiber is with the Dipartimento di Elettronica, Informazione e Bioingegneria,Politecnico di Milano, Via Ponzio, 34/5 - 20133 Milan, Italy - fabio.schreiber@polimi.it -
M.E. Valcher is with the Department of Information Engineering, University of Padova,  via Gradenigo 6B - 35131 Padova, Italy - meme@dei.unipd.it}}
\maketitle

\begin {abstract}
The emergence of Context-aware systems in the domains of autonomic, monitoring, and safety-critical applications asks for the definition of methods to  formally assess their correctness and dependability properties. Many of these properties are common to Automatic Control systems, a field that developed well established analysis and design techniques to formalize and investigate them. 
In this paper, we use Boolean Control Networks, to discuss some properties of a feedback Context-aware system in a case study based on a healthcare management example.\\

\end {abstract}


\section{Introduction and Related Works} \label{Intro}

The growing complexity of modern software systems stimulated the use of com\-ponent-based approaches and the enforcement of the separation of concerns \cite{Djoudi2016}. In Context-aware computing the separation is made between the functions the system is built for, that can change in time owing to different conditions, and the context into which the system must operate, which sets the current environmental situation.

Among the most widely used definitions of Context, and of Context-aware Computing, those proposed by A. Dey \cite{Dey2001} state: {\sl{``Context is any information that can be used to characterize the situation of an entity. An entity is a person, place, or object that is considered relevant to the interaction between a user and an application, including the user and applications themselves."}} and {\sl{``A system is Context-aware if it uses context to provide relevant information and/or services to the user, where relevancy depends on the user�s task."}}

Context-aware applications have been used to: (i) tailor the set of application-relevant data, (ii) increase the precision of information retrieval, (iii) discover services, (iv) build smart environments, et cetera, and different models of the context have been proposed \cite{Bolchini2009,Bolchini2007}. \\ 

However, new application domains such as self-adapting systems \cite{Schreiber2017}, safety critical applications, autonomous vehicle design and manufacturing, disaster prevention, or healthcare management, require a very high level of dependability  that can only be achieved by formally determining their behaviors. To this end, Bigraphs and Model-checking \cite{Djoudi2016,Cherfia2014} approaches have been proposed. In \cite{Padovitz2004,Padovitz2005} Padovitz et Al. consider a state-space approach to describe the \emph{situation} dimension and to determine the likelihood of transitions between  \emph{situation subspaces}, all other Context dimensions remaining constant; the likelihood of the transition is evaluated by assuming notions analogous to those of velocity and acceleration in mechanical systems.\\

Properties as:

\begin{itemize}
\item the existence of \textit{stable equilibrium points}; 
\item the absence of \textit{undesired} oscillations (limit cycles); 
\item   \textit{observability} - the measure of how well internal states of a system can be inferred from knowledge of its external outputs (and, possibly, of its corresponding inputs); 
\item   \textit{controllability} - the ability of an external input (the vector of control variables) to drive the internal state of a system from any initial state to any other final state in a finite time interval;  
\item   \textit{reconstructibility} - when the knowledge of the input and output vectors in a discrete time interval allows to uniquely determine the system final state;
\end{itemize}

are but some of the features, together with \textit{fault detection}, that allow to guarantee the expected and safe operation of a system. 

Systems theorists are well acquainted with the techniques to prove such properties and in \cite{Diao2005} the authors explore {\sl{``... the extent to which control theory can provide an architectural and analytic foundation for building self-managing systems ..."}}. However, control systems are typically described by means of differential equations and by Matrix Algebra, while Context-aware systems are digital and mostly based on Logics. 

Through the introduction of Boolean Control Networks (BCN) and of the semitensor product of matrices, the representative equations of a logic system have been converted into an equivalent algebraic form \cite{Cheng2010b,Cheng2010a}, and solutions to problems such as controllability, observability, stability and reconstructibility have been proposed \cite{Cheng2009,EF_MEV_BCN_obs2012,Fova2016,Zhang_obs}. \\ 

In a previous paper \cite{SchreiberValcher2019}, we proposed and analyzed a BCN model of an open loop Context-aware early-warning hydrogeological system for which we proved: i) the existence of equilibrium points corresponding to constant inputs; ii) the absence of limit cycles; iii) its reconstructibility; iv) the possibility of detecting stuck-in-faults.  

In this paper we consider a multiple feedback loops system 
as it naturally arises when   modeling the evolution of a patient's health status, subjected to medical therapies, whose vital parameters are, in turn,  used as inputs to update the   therapies 
to be administered to the patient. The model  provides the mathematical formalization of a possible algorithm, running on
 the mobile device of a nurse in a hospital,  aiming at providing him/her  with all and only the information on the therapies the patients in his/her ward are to be given.
To focus on the ideas and on the modeling techniques, rather than on the   Boolean math, we 
have chosen to address the model structure and properties without assigning specific numerical values to the logic matrices involved in the system description. Thus we have derived general results that can be tailored to the specific needs
and choices of the illness forms, therapies and vital parameters. We believe that this is the power of the proposed modelling approach: its flexibility and generality. \\
Finally, we provide here a deterministic model of the patient's health evolution, that represents  the evolution of the average case of a patient affected by a specific form of illness. Accordingly, we are giving certain interpretations to the patient's symptoms, as captured by the values of his/her vital parameters, and, based on them, we apply well-settled medical protocols to prescribe therapies and locations where such therapies need to be administered.
A probabilistic model of the patient's reaction to therapies, that also keeps into account the probabilistic correlation between actual health status and the measured values of his/her vital parameters,  requires the use of Probabilistic Boolean Networks and will be the subject of our future research.\\
The paper is organised as follows.  In Section \ref{case} we describe the case study; we model the Context as well as the functional system as Boolean Control Networks, as explained in Section \ref{model}. In Section \ref{reallife} the mathematical formalisation of real life requirements is presented and Section \ref{last} brings some conclusive remarks.

\section{The Case Study} \label{case}

 A hospital keeps a Database that stores all the data relevant both to the patients and to the administrative, medical, and assistance employees. The work of a nurse is guided by an application on his/her mobile device. 
 The App assists the nurse in his/her routine work and is fed by the physician's diagnostic and prescription activities.
\\
Each patient is provided with healthcare wearable sensors  measuring the variables that characterize his/her Medical Status, in our example: the body temperature (\textit{bt}), the blood pressure (\textit{bp}), and the heartbeat frequency (\textit{hf}) \cite{wearable2013, wearable2020}. For ease of representation, all of these variables are discretized and  take values in the finite set\ $S=\{low, medium, high\}$.\\

\begin{figure*}[h]
\vspace{-0.8 cm}
	\includegraphics[scale=.5]{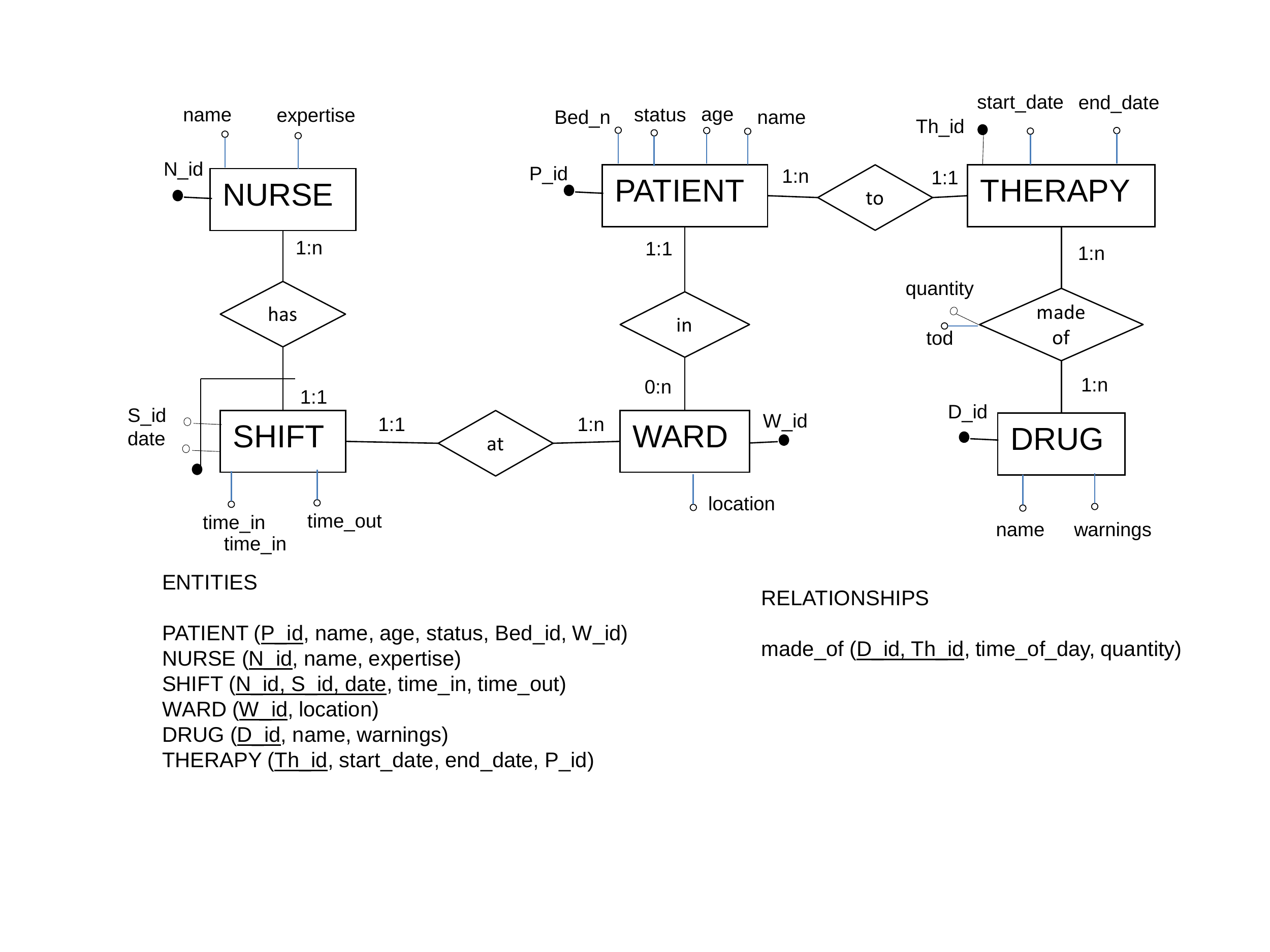}
	\vspace{-2.1 cm}	
	\caption{The hospital Database}
	\label{fig:DB}
\end{figure*}

\begin{figure*}[h]
\vspace{-1.0 cm}
\hspace{-1cm}
		\includegraphics[scale=.5]{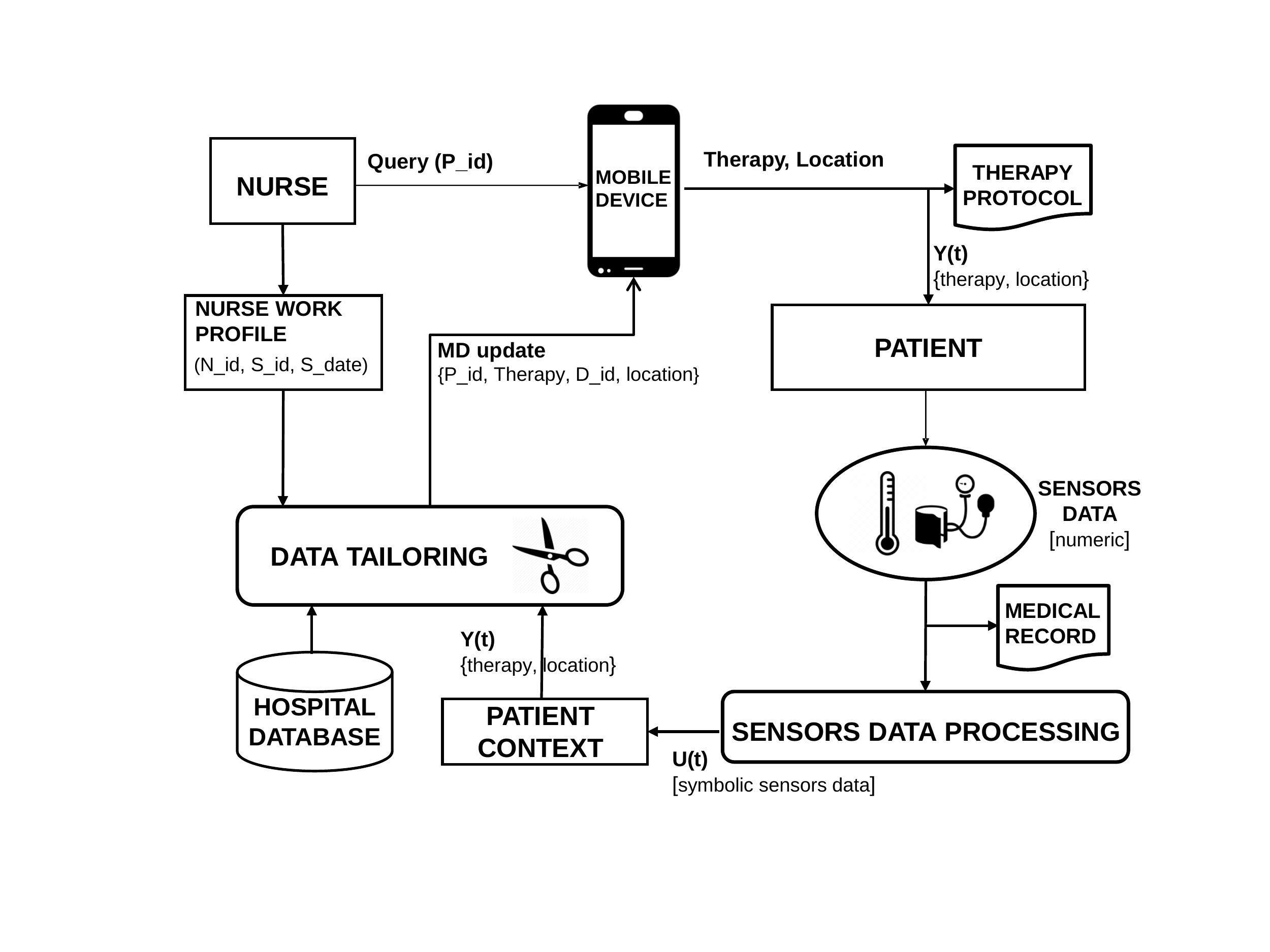}
	\vspace{-1.4 cm}	
	\caption{The tailoring process}
	\label{fig:tailor}
\end{figure*}

Figure \ref{fig:DB} shows a portion of the schema of the hospital Database, which must be dynamically tailored in order to store, on the mobile device of each nurse in a shift, \emph{all and only the treatments each patient in his/her ward is to be given in that shift};  treatments are defined in the Therapy Protocol adopted for the diagnosed illness. The numerical values coming from the sensors - registered in the Medical Record -  are converted into their symbolic aggregate counterparts $ \{low, medium, high\}$ in the Sensors Data Processing block and affect the Estimated Patient Status, which  can take five values: Healthy (\textit{H}), Convalescent (\textit{C}),  Under Observation (\textit{UO}), Ill (\textit{I}), and Life Critical (\textit{LC}).  The Estimated Patient Status determines the physician's decision on both the therapy and  the patient's location -  at home (\textit{h}), in hospital ward (\textit{hw}), in an intensive care unit (\textit{icu}). Clearly,  the prescribed therapies depend also on the current location and on the location that is recommended for the patient. For instance, some therapies can be given in a hospital \textit{icu} or in a \textit{ward}, but cannot be given at \textit{home}.
On the other hand, the medical context can require a relocation of the patient.  Figure \ref{fig:tailor} shows the overall tailoring process.
\\
Thus, Data tailoring is made on the basis of two different criteria:

\begin{itemize}
\item the work profile of the nurse, which is used to select all and only the patients he/she must attend; it is downloaded at the  beginning of the shift and is not affected by external events (Listing \ref{lst:list1});
\item the medical status of the patient, which dynamically requires different treatments.
\end{itemize}

{\tt
\begin{lstlisting}[frame=single,label=lst:list1,caption={Tailoring the Nurse work profile}]
select P_id,bed_n
from nurse,shift,ward,patient
where  N_id="A" AND S_id="X" AND S_date="yy/mm/dd" 
\end{lstlisting}
\label{list1}}

\begin{figure*}[h]
\vspace{-1.5 cm}
		\includegraphics[scale=.5]{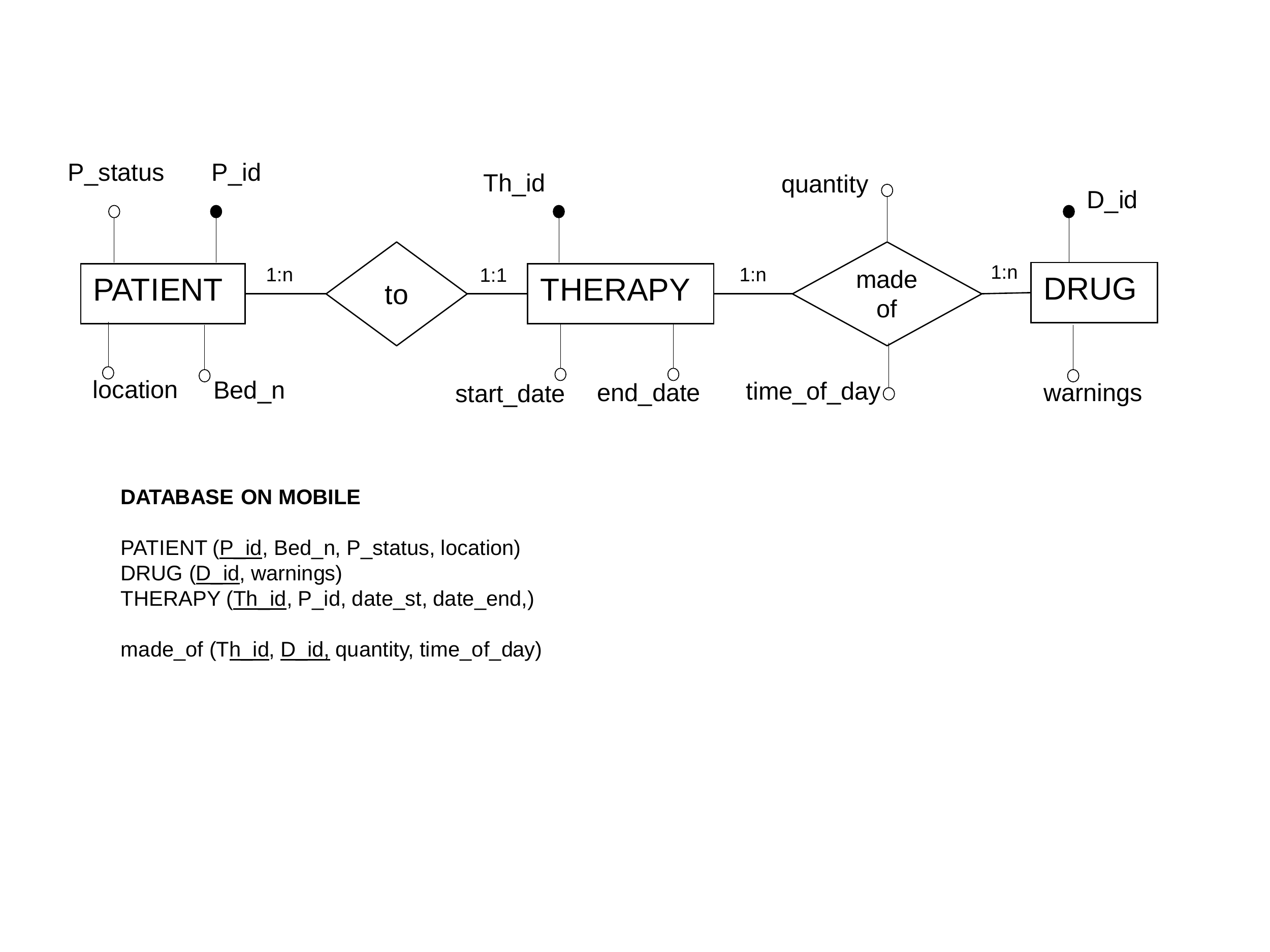}
	\vspace{-3.5 cm}	
	\caption{The tailored Database}
	\label{fig:DBmob}
\end{figure*}

 The query on the nurse's mobile device is shown in  Listing \ref{lst:list2}. For the purpose of this work, in the following, we focus only on the medical and not on the administrative issues.  The schema of the tailored data, stored on the mobile device, is shown in Figure \ref{fig:DBmob}.\\

{\tt
\begin{lstlisting}[frame=single,label=lst:list2,caption={Querying the nurse's device}]
select D_id, quantity, location
from patient,therapy,made_of,D_id
where  P_id="pp"  AND time_of_day="hh:mm"
\end{lstlisting}
\label{list2}}

There are $3{^3}=27$ possible combinations (triples) of the sensors symbolic data - synthetically resumed in Table 1. %\ref{tab:tab1}.

\begin{table}[ht]\label{tab:tab1}
\caption{Possible input combinations}
\centering 
\scriptsize
\begin{tabular}{p{0.4cm}|p{1.6cm}|p{1.6cm}|p{1.6cm}}
  
 \hline\hline %inserts double horizontal lines
  & bt & bp & hf \\ [0.5ex] \hline\hline 
1 & low & low & low \\ \hline 
2 & low & low & mid \\ \hline
3 & low & low & high \\ \hline
4 & low & mid & low \\ \hline
5 & low & mid & mid \\ \hline
6 & low & mid& high \\ \hline
7 & low & high & low \\ \hline
8 & low & high & mid \\ \hline
9 & low & high & high \\ \hline
10 & low & low & low \\ \hline
11 & low & low & mid \\ \hline
12 & low & low & high \\ \hline
13 & mid & mid & low \\ \hline
14 & mid & mid & mid \\ \hline
15 & mid & mid & high \\ \hline
16 & mid & high & low \\ \hline
17 & mid & high & mid \\ \hline
18 & mid & high & high \\ \hline
19 & high & low & low \\ \hline
20 & high & low & mid \\ \hline
21 & high & low & high \\ \hline
22 & high & mid & low \\ \hline
23 & high & mid & mid \\ \hline
24 & high & mid & high \\ \hline
25 & high & high & low \\ \hline
26 & high & high & mid \\ \hline
27 & high & high & high \\ \hline

\end{tabular}
\end{table}

 As detailed in Section \ref{model}, the Patient Context is constituted by a set of variables 
and it determines the therapy to be given (e.g., the drugs, their amount and timing). The treatment  should change the actual Patient Status -  thus changing the sensors output - and, possibly, could require a repositioning of the patient to a different location, so determining  feedback loops.  Furthermore, even if the model is general and we do not enter in the diagnose-prescription issues, we suppose that \textit{the therapies are effective and that the patient will be eventually dismissed}.\\

\begin{center}
\begin{figure*}[ht]
\hspace{-1cm}
		\includegraphics[scale=.55]{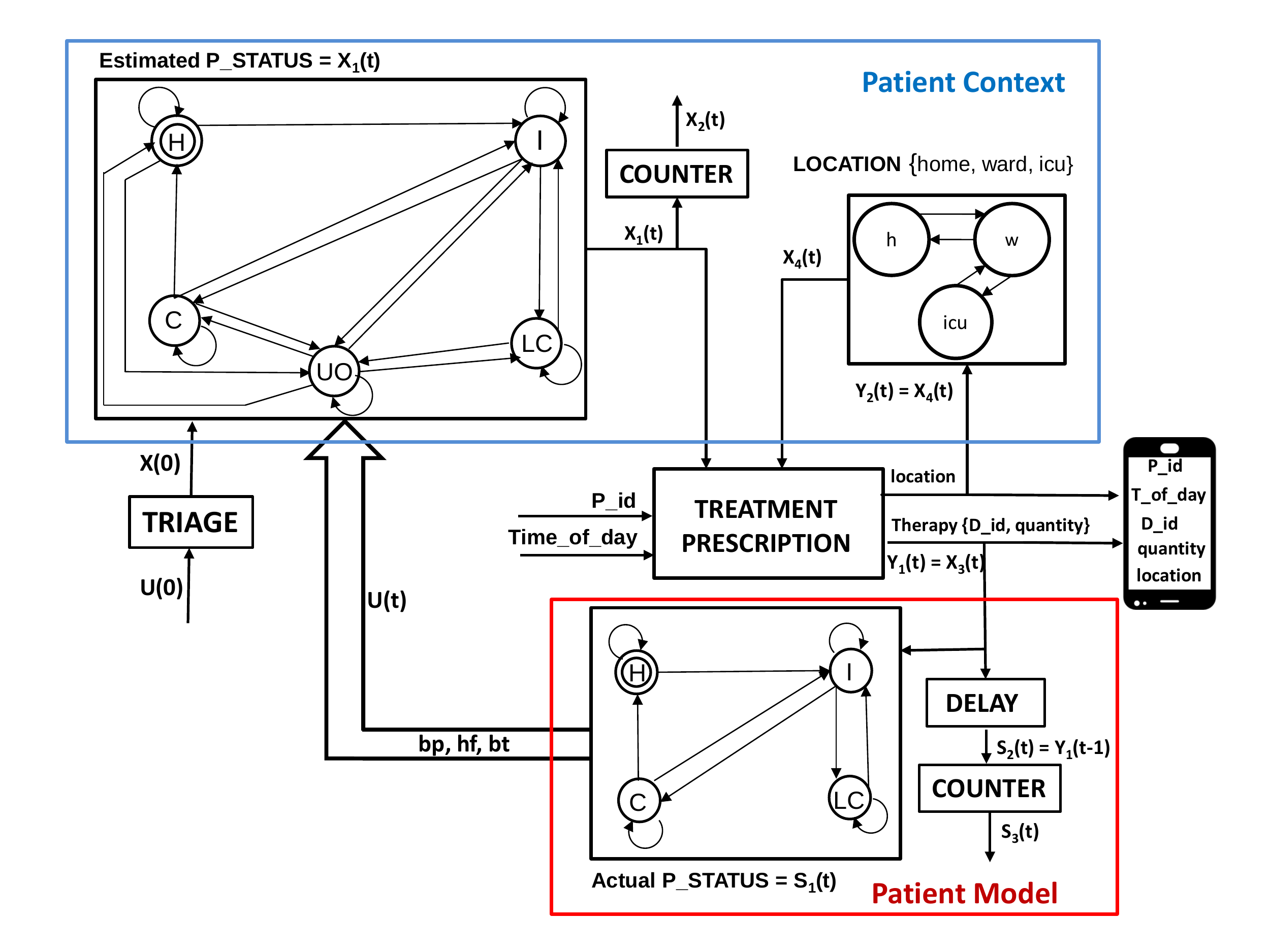}
	\vspace{-1 cm}	
	\caption{The system structure}
	\label{fig:status}
\end{figure*}%}
\end{center}

In Figure \ref{fig:status} the global system structure is represented showing the Moore state diagrams of the Estimated and the Actual Patient Status, and of the Location respectively, as it will be detailed in Section \ref{model}. \\

\section{The System Model}\label{model}

 Before proceeding, we introduce some minimal notions about the left semi-tensor product and the algebraic representations of Boolean Networks and Boolean Control Networks. The interested reader is referred to 
\cite{BCNCheng} for a general introduction to this class of models and to their fundamental  properties. Additional references for the specific properties and results we will use in the paper will be introduced in the following.\\
We consider Boolean vectors and matrices, taking values in ${\mathcal B}  = \{0,1\}$, with the usual
logical operations (And, Or, and  Negation). $\delta^i_{k}$   denotes the $i$th canonical vector of size $k$, 
namely the $i$th column of the $k$-dimensional identity matrix $I_k$. ${\mathcal L}_{k}$ is
the set of  all  $k$-dimensional canonical vectors, and ${\mathcal L}_{k\times n}\subset {\mathcal B}^{k\times n}$  the set of all $k\times n$ {\em logical matrices}, namely $k\times n$ matrices whose  columns are   canonical vectors of size $k$. 

 Boolean variables $X\in {\mathcal B}$ and vectors ${\bf x}\in {\mathcal L}_2$ are related by  a bijective correspondence, defined by the identity
 $${\bf x} = \left[\begin{matrix} X\cr \overline{X}\end{matrix}\right].$$
The {\em (left) semi-tensor product} $\ltimes$ between matrices (in particular, vectors) is defined as follows \cite{BCNCheng}:
given $L_1\in {\mathcal L}_{r_1 \times c_1}$ and $L_2\in {\mathcal L}_{r_2\times c_2}$, we set 
$$L_1\ltimes L_2 := (L_1 \otimes I_{T/c_1})(L_2 \otimes I_{T/r_2}),
\quad {\rm with}\quad T:= {\rm l.c.m.}\{c_1,r_2\}.$$
The semi-tensor product generalizes the standard matrix product,  meaning that when $c_1=r_2$, then
$L_1 \ltimes L_2=L_1L_2$.
In particular, when ${\bf x}_1\in {\mathcal L}_{r_1}$ and ${\bf x}_2\in {\mathcal L}_{r_2}$, we have
${\bf x}_1 \ltimes {\bf x}_2\in  {\mathcal L}_{r_1r_2}.$
 By resorting to the semi-tensor product,   the previous   correspondence extends to a bijective correspondence   \cite{BCNCheng} between ${\mathcal B}^n$ and ${\mathcal L}_{2^n}$. Indeed, given $X= \left[\begin{matrix}X_1 & X_2 & \dots & X_n\end{matrix}\right]^\top\in {\mathcal B}^n$,
one can set  
$${\bf x} := \left[\begin{matrix}X_1\cr \overline{X}_1\end{matrix}\right] \ltimes \left[\begin{matrix}X_2\cr \overline{X}_2\end{matrix}\right]\ltimes \dots \ltimes \left[\begin{matrix}X_n\cr \overline{X}_n\end{matrix}\right],$$
which  corresponds to
$${\bf x}= \left[\begin{matrix}X_1X_2\dots X_{n-1} X_n & X_1X_2\dots X_{n-1} \ \overline{X}_n &  X_1X_2 \dots \overline{X}_{n-1} X_n & \dots &
\overline{X}_1\overline{X}_2\dots \overline{X}_{n-1} \overline{X}_n\end{matrix}\right]^\top.$$
A {\em Boolean Control Network}  (BCN) is  a logic state-space model taking the form:
\be
\begin{array}{rcl}
X(t+1) &=& f(X(t),U(t)), \cr
Y(t)&=& h(X(t),U(t)), \qquad t \in \mathbb{Z}_+,
\end{array}
\label{BCNL}
\ee
where $X(t)$, $U(t)$  and $Y(t)$ are
 the $n$-dimensional state variable,  the $m$-dimensional input variable and the $p$-dimensional output variable at   time $t$, taking values in ${\mathcal B}^n$,    ${\mathcal B}^m$ and ${\mathcal B}^p$, respectively.
 $f$ and $h$ are  logic functions, i.e. $f: {\mathcal B}^n \times {\mathcal B}^m \rightarrow {\mathcal B}^n$, while 
 $h: {\mathcal B}^n \times {\mathcal B}^m \rightarrow {\mathcal B}^p$.
By making use of  the   semi-tensor product $\ltimes$, the BCN (\ref{BCNL}) can be equivalently represented as \cite{BCNCheng}
\be
\begin{array}{rcl}
{\bf x}(t+1)  &=& L \ltimes {\bf u}(t)\ltimes {\bf x}(t), \cr
{\bf y}(t) &=& H \ltimes {\bf u}(t) \ltimes {\bf x}(t), \qquad t \in \mathbb{Z}_+,
\end{array}
\label{BCNA}
\ee
where $L \in \mathcal{L}_{N \times NM}$ and $H \in \mathcal{L}_{P\times NM}$, $N := 2^n, M := 2^m$ and $P :=2^p$. This is called the {\em algebraic expression} of the BCN. The matrix $L$ can be partitioned into $M$ square blocks of size $N$, namely as
$$L = \begin{bmatrix} L_1 & L_2 & \dots & L_{M}\end{bmatrix}.$$
For every $i\in \{1,2,\dots, M\}$, the matrix $L_i\in {\mathcal L}_{N\times N}$ represents the logic matrix that relates ${\bf x}(t+1)$ to ${\bf x}(t)$, when ${\bf u}(t)=\delta^i_{N}$, namely
$${\bf u}(t)=\delta^i_{M} \ \Rightarrow \ {\bf x}(t+1)= L_i {\bf x}(t).$$
In the special case when the logic system has no input, its algebraic expression becomes
\be
\begin{array}{rcl}
{\bf x}(t+1)  &=& L   {\bf x}(t), \cr
{\bf y}(t) &=& H  {\bf x}(t), \qquad t \in \mathbb{Z}_+,
\end{array}
\label{BNA}
\ee
and it is called {\em Boolean Network}.\\
It is easy to realize that the previous algebraic expressions \eqref{BCNA} and \eqref{BNA} can be adopted to represent any state-space model in which the state, input and output variables take values in finite sets, and hence the sizes of the state, input and output vectors  $N, M$ and $P$ need  not   be powers of $2$. When so, oftentimes BCNs and BNs are called multi-valued Control Networks \cite{BCNCheng}. With an abuse of terminology, in this paper we will always refer to them  as BCNs and BNs. Also, in the following capital letters will be used to denote the original vectors/variables, taking values in finite sets, and the same 
lowercase letters will be used to denote the corresponding canonical vectors.\\

With these preliminary definitions and notations, we are now in a position to introduce the BCN models for our case study.

\subsection{The Patient Context Model}

Let us first consider the ``Patient Context" model. According to Table 1, %\ref{tab:tab1}, 
we  assume as input vector
the 3-dimensional vector $U(t)$, where 
\begin{itemize}
\item $U_1(t)$ denotes the (\textit{low, medium or high}) value of  the body temperature (\textit{bt}) at time $t$;
\item $U_2(t)$ denotes the (\textit{low, medium or high}) value of  the body pressure (\textit{bp}) at time $t$;
\item $U_3(t)$ denotes the (\textit{low, medium or high}) value of  the heart frequency (\textit{hf}) at time $t$.
\end{itemize}
The corresponding canonical vector, $u(t)$, therefore belongs to ${\mathcal L}_{27}$, 
since each of the variables $U_i(\cdot), i=1,2,3,$ can take three distinct values (see Table 1). %\ref{tab:tab1}).
\medskip

The state variable $X(t)$ is a 4-dimensional vector, where
\begin{itemize}
\item $X_1(t)$ denotes the Estimated Patient Status (in other words, the Diagnosis) at time $t$ with respect to a specific form of illness: it  takes   values in the set $\{H,C,UO,I,$ $LC\}$;
\item $X_2(t)$ represents a counter variable, that keeps tracks of how many  consecutive times up to time $t$ the Estimated Patient Status has remained invariant. In other words, $X_2(t)=m$ if $X_1(t)=X_1(t-1)= \dots = X_1(t-m+1)$, but  $X_1(t-m+1)\ne X_1(t-m)$. In order to ensure that  $X_2$ takes values in a finite set, and for the sake of simplicity\footnote{All the numbers used in this context are, of course, arbitrary and  meant to purely exemplify how to design the algorithm and to convert it into a BCN.}, 
we assume that we   keep track until $X_2(\cdot)$ reaches the value $3$, and then we stop. This amounts to saying that  $X_2(t)$ belongs to $\{1,2,\ge 3\}$;
\item $X_3(t)$ is 
the prescribed therapy at time  $t$, belonging to a finite set, say $\{Th0, Th1,$ $ \dots,Th5\}$, where $Th0$ means that the patient does not receive any  drug;
\item $X_4(t)$ is  the prescribed location   (\textit{home, ward, icu}) where the patient will get the therapy at time $t$.
\end{itemize}
The corresponding canonical representation, $x(t)$,  under the previous assumptions  will belong to ${\mathcal L}_{270}$, since $270=5\times 3\times 6\times 3$.
\medskip

Finally, we assume as output of the Patient Context the 2-dimensional vector
 $Y(t)$,  where
\begin{itemize}
\item $Y_1(t)$  is 
the prescribed therapy at time  $t$;
\item $Y_2(t)$  is  the prescribed location (\textit{home, ward, icu}) where the patient will get the therapy at time $t$.
\end{itemize}
Clearly, $Y_1(t)=X_3(t)$ and $Y_2(t)=X_4(t)$. Moreover, the canonical representation of $Y(t)$, $y(t)$, belongs to  ${\mathcal L}_{18},$ since 18 is the number of possible combinations of therapies and locations. Note, however, that the set of possible outputs can be significantly reduced: for instance, the location \textit{home} is compatible only with the choice to dismiss the patient, after considering his/her health status, and with prescribed  
therapies such as \textit{Th0} (no drugs) or a light therapy (say, \textit{Th1}). At the same time certain therapies can be administered only when the patient is in the \textit{icu}. So, one may reasonably assume that a good number of the 18 output values are not realistic and hence can be removed, thus reducing the size of $y(\cdot)$.   \\
It is worthwhile to introduce a few comments about the initial state $X(0)$ (or its canonical representation $x(0)$)
and about the update of the state variables $X_i(t),i=1,2,3,4$.
The initial state can be regarded as the result of the triage process: when patients are admitted to the  \textit{Emergency Room (ER)}, a preliminary diagnosis is made based on the three measures $U_1(0), U_2(0)$ and $U_3(0)$, since there may be no previous history of the patient and the hospital admission requires a fast evaluation of the medical conditions of the patient. So, $X_1(0)$ may be a static function of $U(0)$. $X_3(0)$ is automatically set to $Th0$, while  $X_2(0)$ is  set to $1$ and $X_4(0)$ to \textit{home}. 
\medskip

We note that $X_1(t+1)$ is naturally expressed as a logic function of $X_1(t), X_2(t),$ $X_3(t), X_4(t)$ and $U(t)$, say $X_1(t+1)=f_1(X(t),U(t))$.
On the other hand, $X_2(t+1)$ naturally depends on $X_2(t), X_1(t)$ and $X_1(t+1)$, and, since we have just pointed out that $X_1(t+1)=f_1(X(t),U(t))$, we can in turn express $X_2(t+1)$ as $X_2(t+1)=f_2(X(t),U(t))$.
Similarly, $X_3(t+1)$ and $X_4(t+1)$ are functions of $X_1(t+1), X_2(t+1),$ $X_3(t), X_4(t)$ and $U(t)$, and hence can be expressed, in turn, as functions of $X_1(t), X_2(t),$ $X_3(t), X_4(t)$ and $U(t)$.\\On the other hand, as we previously remarked, 
$Y_1(t)=X_3(t)$ and $Y_2(t)=X_4(t)$.
This implies that
$$X(t+1)=f(X(t),U(t)),$$
while
$$Y(t) = \begin{bmatrix} X_3(t)\cr X_4(t)\end{bmatrix},$$
and hence
\begin{eqnarray*}
x(t+1) &=& L \ltimes u(t) \ltimes x(t),  \\
y(t) &=& M x(t),   \qquad t\in {\mathbb Z}_+, 
\end{eqnarray*}

for suitable choices of the logical matrices $L\in {\mathcal L}_{270\times (27 \cdot 270)}$ and $M\in {\mathcal L}_{18\times 270}.$
\medskip

\subsection{The Patient Model}

At this point we consider the Patient Model. A reasonable choice of the Patient state variables is the following one:

\begin{itemize}
\item
$S_1(t)$ represents the actual Patient Status that takes values in the set $\{H,C,I,LC\}$. Note that this is a proper subset of the set where the Estimated Patient Status takes values, since of course the value \textit{UO} in this case does not make sense.
\item
$S_2(t)$ represents the therapy that has been prescribed at time $t-1$, and hence it coincides with $Y_1(t-1)$.\\
\item $S_3(t)$ is a counter variable that keeps track of how many consecutive times   up to time $t$ the therapy has remained invariant. In other words, $S_3(t)=m$ if $S_2(t)=S_2(t-1)= \dots = S_2(t-m+1)$, but  $S_2(t-m+1)\ne S_2(t-m)$. 
Also in this case we put a bound on $m$ and assume that
 $S_3(t)$ belongs to $\{1,2,\ge 3\}$.\\
\item Finally, $S_4(t)$ is the vector collecting the measures of the vital parameters at time $t-1$, namely $S_4(t)=U(t-1)$.
\end{itemize}
For the Patient Model, the natural input is $Y(t)$ (in fact, $Y_1(t)$ could be regarded as enough), while the output is $U(t)$.
Since $U(t)$ is the patient's vital parameters at time $t$, it is reasonable to assume that these measures depend on their own values at time $t-1$ (and hence  on $S_4(t)$), on the Patient Status $S_1(t)$, the given therapy at time $t-1$, $S_2(t)$, (indeed it is not realistic to assume that the effect of  the therapy is instantaneous) and on the duration of the therapy,   namely on $S_3(t)$.
\\
With reasonings similar to the ones adopted for the Patient Context model, we can claim that the  Patient Model is described by the logic equations
\begin{eqnarray*}
S(t+1)&=&f_p(S(t),Y(t),U(t)),\\
U(t) &=& h_p(S(t)),
\end{eqnarray*}
and hence by the BCN
\begin{eqnarray*}
s(t+1) &=& F \ltimes y(t)\times u(t) \ltimes s(t),  \\
u(t) &=& H s(t),   \qquad t\in {\mathbb Z}_+,  
\end{eqnarray*}
for suitable choices of the logical matrices $F\in {\mathcal L}_{1944\times (18\cdot 27 \cdot 1944)}$ and $H\in {\mathcal L}_{27\times 1944},$ since $1944= 4\cdot 6\cdot 3\cdot 27$.
\\
So, to summarize, we have the   following two models:
\begin{eqnarray}
x(t+1) &=& L \ltimes u(t) \ltimes x(t), \label{pc1} \\
y(t) &=& M x(t),   \qquad t\in {\mathbb Z}_+, \label{pc2} 
\end{eqnarray}
and
\begin{eqnarray}
s(t+1) &=& F \ltimes y(t)\times u(t) \ltimes s(t), \label{p1} \\
u(t) &=& H s(t),   \qquad t\in {\mathbb Z}_+, \label{p2} 
\end{eqnarray}

 In the following, for the sake of simplicity, we will use the following notation:
 $270= {\rm dim}\ x =:N_x$, $1944= {\rm dim}\ s =: N_s$, $27={\rm dim}\ u= N_u$ and $18= {\rm dim}\ y =: N_y$.\\
If we replace \eqref{p2} and  \eqref{pc2} in \eqref{p1}, and keep into account that
$$s(t)\ltimes s(t) = \Phi s(t),$$
where $\Phi\in {\mathcal L}_{N_s^2\times N_s}$ is a logical matrix known as {\em power-reducing matrix} \cite{BCNCheng},
then  \eqref{p1}
becomes
\begin{equation}
s(t+1)  = F \ltimes  M   \ltimes x(t) \ltimes H\ltimes  \Phi \ltimes s(t). \label{p1a}
\end{equation}
At the same time, we can swap, namely reverse the order of, the vector $x(t)$ and the vector $H\ltimes  \Phi \ltimes s(t)$ by resorting to the {\em swap matrix} $W$ of suitable size   \cite{BCNCheng}, %{\color{red} corretto???} 
thus obtaining
\begin{eqnarray}
s(t+1)  &=& F \ltimes  M   \ltimes  W \ltimes H\ltimes  \Phi \ltimes s(t) \ltimes  x(t) \nonumber \\
&=& A (s(t) \ltimes x(t)), \label{P}
\end{eqnarray}
where 
$$A :=  F \ltimes  M   \ltimes  W \ltimes H\ltimes  \Phi  \in {\mathcal L}_{N_s \times N_s N_x}.$$
Similarly,
if we replace \eqref{p2}  in \eqref{pc1} we get:
\begin{eqnarray}
x(t+1) &=& L \ltimes H \ltimes  s(t) \ltimes x(t)\nonumber \\
&=& B  (s(t) \ltimes x(t)),\label{PC} \end{eqnarray}
where
$$B := L \ltimes H\in {\mathcal L}_{N_x \times N_s N_x}.$$
\\
Now, the overall model, keeping into account both the Patient Context and the Patient Model,
becomes:
\begin{eqnarray}
s(t+1)  &=& A \ltimes s(t) \ltimes x(t),\\
x(t+1) &=& B \ltimes s(t)\ltimes x(t).
\end{eqnarray}
If we introduce the status of the overall system
$$v(t) := s(t)\ltimes x(t) \in {\mathcal L}_{N_sN_x},$$
we get 
$$
v(t+1) = (A \ltimes v(t)) \ltimes (B \ltimes v(t)).$$
It is a matter of elementary calculations to verify that
  once we denote by $a_i$ the $i$-th column of $A$ and by $b_j$ the $j$-th column of $B$, the previous equation can be equivalently rewritten as
\begin{equation}
v(t+1) =  W \ltimes v(t),
\label{stato_v}
\end{equation}
where
$$W := \begin{bmatrix} a_1 \ltimes b_1 & a_2\ltimes b_2 & \dots & a_{N_sN_x}\ltimes b_{N_sN_x}\end{bmatrix}
\in {\mathcal L}_{(N_sN_x)\times (N_sN_x)}.$$
In addition, one can assume as system output
$$y(t)= M x(t)$$
that can be rewritten as
\be
y(t)= \Psi v(t),
\label{output_v}
\ee
where 
$$\Psi := \begin{bmatrix} M & M & \dots & M\end{bmatrix} \in {\mathcal L}_{N_y\times N_sN_x}.$$
So, equations \eqref{stato_v} and \eqref{output_v} together describe a BN that models the overall closed-loop system.
\medskip

\section{Real life requirements and their mathematical formalization}\label{reallife}

In this section we investigate the properties of the overall system, obtained by the feedback connection of the Patient Context and of the Patient Model, namely the BN \eqref{stato_v}-\eqref{output_v}.
\\
As stated in Section \ref{Intro}, we aim to provide general ideas about the mathematical properties of the system that have a clear practical relevance in this context, rather than to check those properties for a specific choice of the logical matrices involved in the system description. Thus, we shall not provide numerical values for the quadruple of logical matrices $(L,M,F,H)$, but we shall show how to reduce our specific feedback system (or parts of it) to standard set-ups for which these properties have already been investigated.

\subsection{Identifiability of the Patient Status}

A first question that is meaningful to pose is whether the Patient Model is a  good one, namely it will lead to a correct functioning of the overall system.
In order to clarify what we mean when posing this question, we first need to better explain  the perspective we have taken in modelling  the patient.
We have assumed that the patient is in a certain medical condition with respect to a specific medical problem. 
 So, the  diagnosis pertains only to the level/seriousness of the patient's health condition, and not to the specific cause of the illness.
  Such a medical condition is unknown to the  nurse, but of course it is the reason why the patient's vital parameters (\textit{bp, bt, hf}), namely the patient's output $U(t)$, take certain values. The medical status is of course affected by the therapy $Y$ and can be associated with different values of $U$, so the output measure $U(t)$ at time $t$ together with the therapy $Y(t)$ (or  $Y_1(t)$) do not allow to uniquely determine $S(t)$. In addition, some therapies may need some time to become effective (which is the reason why we introduced the state variable $S_2(t)$).\\
On the other hand, a good (deterministic) model of the patient\footnote{As previously mentioned, we have adopted a deterministic model and assumed that everything works according to statistics and well settled procedures: therapies are designed according to specific protocols and statistically lead to the full recovery of the patient. This is the reason why the possibility that the patient dies is not contemplated.}
necessarily imposes that the measured vital parameters are significant and hence allow physicians to determine the  actual Patient's Status after a finite number of observations.
\\
From a mathematical point of view, this amounts to assuming that \textit{the Patient Model \eqref{p1}-\eqref{p2} is reconstructible}, namely there exists $T\in {\mathbb Z}_+$ such that  the knowledge of the signals 
$u(\cdot)$ and $y(\cdot)$ in $[0,T]$ allows to uniquely determine $s(T).$
Specifically, we have the following definition:\\

\begin{definition}
The BCN \eqref{p1}-\eqref{p2}, with $s(t)\in {\mathcal L}_{N_s},   u(t)\in {\mathcal L}_{N_u}$ and $y(t)\in {\mathcal L}_{N_y}$, is said to be   {\em reconstructible} if there exists $T\in {\mathbb Z}_+$ such that the knowledge of the input and output vectors in the discrete interval $\{0,1,\dots, T\}$ allows to uniquely determine the final state $s(T)$.
\end{definition}
\medskip

It is worth noticing that the BCN \eqref{p1}-\eqref{p2} is different from the standard ones for which the observability and reconstructibility problems have been addressed in the literature (see  \cite{EF_MEV_BCN_obs2012,MM_obs,Zhang_obs}), since this BCN is intrinsically in a closed-loop condition, as the BCN output $u(t)$ affects the state update at time $t+1$.
However, by replacing \eqref{p2} in \eqref{p1}, and by using again the power reducing matrix, we can obtain:
\begin{eqnarray*}
s(t+1) &=& F \ltimes y(t)\times H \ltimes \Phi \ltimes s(t),   \\
u(t) &=& H s(t),   \qquad t\in {\mathbb Z}_+,  
\end{eqnarray*}
which, in turn, can be rewritten as
\begin{eqnarray}
s(t+1) &=& {\mathbb F}  \ltimes y(t) \ltimes s(t), \label{p1_rec} \\
u(t) &=& H s(t),   \qquad t\in {\mathbb Z}_+, \label{p2_rec} 
\end{eqnarray}
where 
$${\mathbb F} := \begin{bmatrix} {\mathbb F}_ 1 & {\mathbb F}_2 & \dots & {\mathbb F}_{N_y}\end{bmatrix}
$$
and $$
{\mathbb F}_i := \begin{bmatrix} f_i\ltimes   (H \ltimes \Phi)\delta^1_{N_s} & \dots & f_i  \ltimes (H \ltimes \Phi)\delta^{N_s}_{N_s}\end{bmatrix},$$
 where we have denoted by $f_i$ the $i$-th column of the matrix $F$.\\
This allows to reduce the reconstructibility problem for this specific BCN to a standard one, for which there are  lots of results and algorithms (see  \cite{EF_MEV_BCN_obs2012,Zhang_obs,ZhangJohansson2020,ZhangLeifeldZhang2019}). \\
Clearly, the matrices $F$ and $H$ must be properly selected in order to guarantee the reconstructibility of the patients' status. This means, in particular, that the vital parameters to measure must be chosen in such a way that they are significant enough to allow to identify the actual medical conditions of the patient.
\medskip

From a less formal viewpoint, it is worth underlying that the reconstructibility problem reduces to the problem of correctly identifying the state variable $s_1(t)$, since the definition of $s_i(t), i=2,3,4,$ allows to immediately deduce that such values can be uniquely determined from the variables $y_1(t)$ and $u(t)$.
 So, one could focus on a lower dimension model expressing $s_1(t+1)$ in terms of $s_i(t), i=1,2,3,4$, $u(t)$ and $y_1(t)$, where $s_i(t), i=2,3,4, u(t)$ and $y_1(t)$ are known, and address the reconstructibility of $s_1(t)$ from $u(t)$, assuming $s_i(t), i=1,2,3,4$,  and $y_1(t)$ as inputs.
\medskip

\subsection{Correct diagnosis}

Of course, once we have ensured that the Patient Model \eqref{p1}-\eqref{p2} is reconstructible, and hence we have properly chosen the vital parameters to measure  in order to identify the Patient Status, the   natural question  arises: Is the Patient Context correctly designed so that after a finite (and possibly small) number of steps $T$, the Patient's Status $s_1(t)$ and the Estimated Patient's Status $x_1(t)$ coincide for every $t\ge T$? This amounts to saying that the protocols to evaluate the Patient Status have been correctly designed.
\\
To formalize this problem, we need to introduce a comparison variable, say $z(t)$.
This variable takes the value $\delta^1_2$ (namely the unitary or YES value) if $s_1(t)=x_1(t)$ and the value
$\delta^2_2$ (namely the zero or NO value) otherwise.
Keeping in mind that $S_1(t)$ takes values in $\{H,C, I,LC\}$ (and hence $s_1(t)\in {\mathcal L}_4$), while 
$X_1(t)$ takes values in $\{H,C,UO,I,$ $LC\}$ (and hence $x_1(t)\in {\mathcal L}_5$),
this leads to
$$z(t)= \begin{bmatrix} C_1 & C_2 & C_3 & C_4 \end{bmatrix} s_1(t)\ltimes x_1(t),$$
where\footnote{ To improve the notation one could sort the set of values of the Estimated Patient State as  follows: $\{H,C, I,LC,UO\}$. In this way, each of the blocks $C_i$  would have the $i$th column equal to $\delta^1_2$ and all the remaining ones equal to $\delta^2_2$.}
$C_i \in {\mathcal L}_{2\times 5}$ for every $i\in [1,4].$ Moreover,\\
$C_1$ is the block whose first column is $\delta^1_2$ while all the others are $\delta^2_2$;\\
$C_2$ is the block whose second column is $\delta^1_2$ while all the others are $\delta^2_2$;\\
$C_3$ is the block whose fourth column is $\delta^1_2$ while all the others are $\delta^2_2$;\\
$C_4$ is the block whose fifth column is $\delta^1_2$ while all the others are $\delta^2_2$.

Clearly, $z(t)$ can also be expressed as a function of $s(t)$ and $x(t)$ and hence as a function of $v(t)$. This leads to
$$z(t) = {\mathbb C} v(t),$$
 for a suitable ${\mathbb C}\in {\mathcal L}_{2\times N_sN_x}$.
So, the problem of understanding whether the system is designed to produce the correct diagnosis can be equivalently translated into the mathematical problem of determining whether for every initial condition, $v(0)$, the output trajectory of the system
\begin{eqnarray}
v(t+1) &=& W v(t) \label{BNlarga}\\
z(t) &=& C v(t)
\end{eqnarray}
eventually takes the value $\delta^1_2$. In other words, we need to ensure that there exists $T\in \mathbb {Z}_+$ such that, for every $v(0)\in {\mathcal L}_{N_sN_x}$, the corresponding output trajectory $z(t), t\in {\mathbb Z}_+,$ satisfies
$z(t)=\delta^1_2$ for every $t\ge T$.\\
  Note that the idea is that once the  seriousness level of the patient's illness has been correctly diagnosed, this information will never been lost, even if the patient's health status will change.\\
Another way of looking at this problem is to define the set of states 
$${\mathcal C}\!\!{\mathcal D} :=\{ v(t)\in {\mathcal L}_{N_sN_x} :  s_1(t)\ltimes x_1(t)\in \{\delta^1_4\ltimes \delta^1_5, \delta^2_4\ltimes \delta^2_5, \delta^3_4\ltimes \delta^4_5, \delta^4_4\ltimes \delta^5_5\}\},$$
that represent all possible situations where the Estimated Patient Status $x_1(t)$ coincides with the 
Patient Status $s_1(t)$ (in other words, ${\mathcal C}\!\!{\mathcal D}$ is the set of correct diagnoses), and to impose that such a set is a global attractor of the system.\\
From a formal point of view, the set 
${\mathcal C}\!\!{\mathcal D}$ is a {\em global attractor of the BN} \eqref{BNlarga}
if there exists $T\ge 0$ such that for every $v(0)\in {\mathcal L}_{N_sN_x \times N_s N_x}$ the corresponding state evolution $v(t), t\in {\mathbb Z}_+$, of the BN \eqref{BNlarga} belongs to ${\mathcal C}\!\!{\mathcal D}$ for every $t\ge T$.\\
This property can be easily checked \cite{BCNCheng,EF_MEV_BCN_Aut} by simply evaluating that all rows
of $W^{N_sN_x}\in {\mathcal L}_{N_sN_x}$, the $N_sN_x$ power of $W$, are zero except for those whose indexes correspond to the canonical vectors in ${\mathcal C}\!\!{\mathcal D}.$

\subsection{Successful therapies}

As previously mentioned, when modeling the evolutions of the Patient Context and of the Patient Model in a deterministic way, we are describing the evolution of the average case of a patient affected by a specific form of illness. Accordingly, we are giving certain interpretations to the patient's symptoms, as captured by the values of his/her vital parameters, and based on them we are applying well-settled medical protocols to prescribe therapies and locations where such therapies need to be administered.
In this context it is clear that death is not contemplated, since this would correspond to assuming that a given medical protocol deterministically leads to the death of the patient and this does not make sense. Similarly, a protocol that deterministically leads  to an equilibrium state where the Patient's Status is $C, I$ or $LC$ is not acceptable. In other words, the only reasonable solution is to have designed the Patient Context in such a way that 1) the Patient Status is eventually H; 2) the Estimated Patient Status is, in turn, H.\\
Note that 1) and 2) correspond to imposing that \textit{the global attractor of the system evolution is a proper subset, say ${\mathcal H}$, of the set
${\mathcal C}\!\!{\mathcal D}$} we previously defined. Specifically,
we define the set ${\mathcal H}$ as follows:
$${\mathcal H} :=\{ v(t)\in {\mathcal L}_{N_sN_x} :  s_1(t)\ltimes x_1(t)= \delta^1_4\ltimes \delta^1_5\},$$
that represent all possible situations where the Estimated Patient Status $x_1(t)$ is healthy and it coincides with the 
Patient Status $s_1(t)$ (in other words, ${\mathcal H}$ is the set of states corresponding to a healthy patient whose health status has been correctly identified), and to impose that such a set is a global attractor of the system\footnote{Note that we are not introducing additional constraints, in particular we are assuming that the vital parameters $u$ of the patient can change within the set of values compatible with the healthy status. Of course, one could further constrain the set ${\mathcal H}$ by assuming that the prescribed therapy is Th0, the patient is at home, and all the counters have reached the saturation level. Even in this case, we may regard as acceptable the existence of a limit cycle, since this would only correspond to oscillations of the values of the state variable $s_4$ within a small set of values that do not raise any concern. Clearly, one may impose also for $s_4$ and hence for $u$ a prescribed desired value, and this would mean asking that the system has a single {\em equilibrium point} (the set ${\mathcal H}$ has cardinality one) which  is a global attractor.}.\\
Also, in this case it is possible to verify whether such a requirement is met by evaluating if  
all rows
of $W^{N_sN_x}\in {\mathcal L}_{N_sN_x}$, the $N_sN_x$ power of $W$, are zero except for those whose indexes correspond to the canonical vectors in ${\mathcal H}.$

\section{Conclusions} \label{last}

In this paper we have used an interesting case study, related to the 
evolution of the health status of a patient,  to illustrate how 
a feedback Context-aware system can be modeled by means of a BCN. Indeed, the patient is subjected to medical therapies and his/her vital parameters are
not only the outcome of the therapies, but also the input based on which therapies are prescribed.
By referring to a simplified deterministic model in terms of BCNs/BNs, we have been able to illustrate how the most natural practical goals that the overall closed-loop system needs to achieve may be formalized, and hence investigated,  by resorting to well-known
System Theory concepts.
Clearly, the given model can be improved and tailored to the specific needs, to account for more complicated algorithms, and more exhaustive sets of data, but the core ideas have already been captured by the current model.
Also, we have addressed what seemed to be the most natural targets in the specific context,  but different or additional properties 
may be investigated, in case the same modeling technique is applied to describe closed-loop Context Aware systems of different nature.\\
The use of a  deterministic model of the patient's health evolution, to plan therapies based on measured vital parameters, represents a first step toward the design of an accurate algorithm to employ in  the 
 mobile device of a nurse.  
A probabilistic model, together with some warning system that advises the nurse of when different
decisions are possible with different confidence levels, and hence there is the need for the immediate supervision 
of a specialist, is the target of  future research.

%\section{References}

\end{document}